\documentstyle[12pt,aps,graphicx]{revtex}
\newcommand{\de}{\partial}
\newcommand{\fr}{\frac}

\newcommand{\g}{\gamma}
\newcommand{\beq}{\begin{equation}}
\newcommand{\eeq}{\end{equation}}

\begin{document}

\input{epsf.tex}
\epsfverbosetrue
\draft

\title{
Multi-filament structures in relativistic self-focusing
}
\author{
F.~Cattani,$^1$ A.~Kim,$^2$ D.~Anderson,$^1$ and M.~Lisak$^1$
}
\address{
$^1$Department of Electromagnetics, Chalmers University of Technology, 412
96 G\"{o}teborg, Sweden\\
$^2$Institute of Applied Physics, Russian Academy of Sciences, 603600
Nizhny Novgorod, Russia
}
\maketitle

\begin{abstract}
A simple model is derived to prove the multi-filament structure of
relativistic self-focusing with  ultra-intense lasers. Exact analytical
solutions describing the transverse structure of waveguide channels with
electron cavitation, for which both the relativistic and ponderomotive
nonlinearities are taken into account, are presented.
\end{abstract}


\section{INTRODUCTION}

Recent development in laser technology has opened up the possibility of
exploring previously  unattainable regimes of laser-plasma interactions,
\cite{perry}. Intensities of the order of $10^{18} W/cm^2$ and higher can
now be achieved, implying that goals like compact sources for x-ray lasers,
\cite{chen}, the fast ignitor concept for inertial confinement fusion, (ICF)
\cite{tabak} and laser-plasma based accelerators, \cite{esarey}, might soon
be within reach. However, a major effort is still required both numerically
and analytically in order to understand the nonlinear phenomena which arise
in the presence of such extremely high intensities of the electromagnetic
radiation. Besides, a good analytical insight is needed in order make
numerical simulations possible and to interpret their results,
\cite{pukhov}.

One of the problems which have received particular attention is the
combined effect of relativistic and striction (ponderomotive)
nonlinearities, which occur in the propagation of superintense laser pulses
through underdense plasmas (i.e., plasmas with $\omega_p < \omega$, where
$\omega_p= (4 \pi n_e e^2/m_e)^{1/2}$ is the plasma frequency and $\omega$
is the laser carrier frequency), \cite{litvak}. This problem is not fully
understood yet and there is  need of a self-consistent analytical
description which does not violate the global charge conservation and
plasma quasineutrality when describing self-focusing and self-channeling,
\cite{feit}.

A common feature in the above mentioned advanced schemes is that transport
of laser radiation over considerable distances, well beyond the diffraction
limit, and without significant energy losses is required. In achieving this
goal, nonlinear self-focusing and self-channeling play an important role.
Under the action of an intense laser pulse, electrons tend to be
redistributed in the transverse direction  as an effect of the
ponderomotive pressure, the self-channeling phenomenon. The subsequent
self-modification of the radial profile of the refractive index is at the
origin of the nonlinear self-focusing and filamentation of the laser pulse.
Nonlinear self-focusing and self-channeling offer a possibility for optical
guiding of laser pulses in underdense pulses as, for instance, the
underdense corona of an ICF target, through the formation of ''hollow
channels'', \cite{chen93}, as experimentally observed by several groups,
\cite{exp,young}.\\
%
%

%
As was shown in \cite{sun}, relativistic self-focusing shows qualitatively new
features
for ultra-intense lasers. The ponderomotive force of super-strong fields expels
electrons thus producing ''vacuum channels'' which guide the
radiation and stable channeling with power higher than the critical one can take
place, \cite{borisov}.
 However, as noticed in \cite{kurki}, analytical
descriptions led to the appearance of negative electron densities. This
problem was solved by setting the channel boundary positions exactly at the
point where the electron density became zero. Feit et al., \cite{feit},
showed that this procedure did not conserve the global charge and proposed
to include the electron temperature effect, which however was not
self-consistently evaluated (the temperature was assumed to be derived from
experimental conditions). Recently, we noticed that for an overdense plasma
a self-consistent description of self-induced transparency is possible
which automatically takes into account global charge conservation through
Poisson's equation, \cite{jetp}. The strong analogies between 1D overdense
and 2D underdense plasmas, allow for an exact analysis of the stationary
stage of electron cavitation due to the joint effects of relativistic and
striction nonlinearities in underdense plasmas. This analysis leads to an
exact analytical description of the transverse structures generated by
relativistic self-focusing and also demonstrates its multi-filament nature.\\
%
%
Depending on the incident power and intensity distribution, several qualitatively
different solutions
may occur. Our
aim is to give an exact analytical description of the stationary stage of the
fundamental configurations. We will show that, if the incident power is relatively low
and the intensity has its maximum  on-axis, the plasma will react by generating a stationary
stage with a single channel acting as an optical guide for the propagating radiation.
If the incident power is increased and the intensity instead has its minimum  on-axis (a higher order
laser mode), then the final stationary stage will display two symmetric channels.
Finally, for even higher incident powers and maximum on-axis incident
intensities, three
channels will be generated and so on, with the critical power for channel formation
depending on the unperturbed plasma density and the wave number of the propagating
radiation.\\
Of particular interest will be the fact that these structures can be
interpreted as the final stationary stage of the filamentation instability,
as shown by the numerical simulations presented in \cite{pukhov}.\\
 In this paper, we introduce, in Section \ref{model},  the
model equations and the approximations we will use to describe electron cavitation
in a two-dimensional underdense
plasma. The commonly presented analysis is discussed in Section \ref{common} and our results for single and multi-channel
structures in 2D, planar geometry  are presented in the following Sections,
following a brief discussion
of the physical mechanisms behind the generation of such structures. Finally,  some
conclusions are summarized in Section \ref{conclusions}.

\section{THE MODEL}\label{model}

Let us consider the propagation of electromagnetic radiation
in a homogeneous plasma. A complete description is based on Maxwell's
equations for the propagating laser radiation plus a model describing the
plasma response. For the problem of interest, an important simplification
comes from the physical context. Considering short pulses with length
$\tau$ such that $\omega_{pe}^{-1}\ll \tau \ll \omega_{pi}^{-1}$, the ion
dynamics can be neglected, \cite{borisov}. Furthermore, we will not be concerned with wake
field generation, \cite{kurki}, since the pulse is long enough to
allow us
to disregard longitudinal charge separation,  \cite{feit}.

Finally, all
thermal effects will be disregarded since, at these high intensities,
electrons are driven to relativistic velocities in just a few optical
cycles and the electron pressure gradient is negligible  compared to the
ponderomotive pressure, \cite{feit96}.\\
These assumptions define the model we are using to
describe our plasma: The ions are considered as an immobile neutralizing
background and the electrons as a cold relativistic fluid. Our set of
self-consistent equations derived from Maxwell's equations and the equation
of motion for the electron component, assuming the Coulomb gauge, reads
\begin{eqnarray} \label{eq1}
\nabla^2 {\bf{A}} - \fr{1}{c^2} \fr{\de^2 {\bf{A}}}{\de
t^2}=\fr{1}{c}\fr{\de}{\de
t}{{\bf \nabla}} \varphi + \fr{4 \pi}{c}N e {\bf{v}}\\
\label{eq2}
\nabla^2 \varphi=4 \pi e (N - N_0)\\
\label{eq3}
m \gamma {\bf{v}} = \fr{e}{c}{\bf{A}} + {\bf{\nabla}} \psi\\
\label{eq4}
\fr{\de \psi}{\de t} = e \varphi - m c^2 (\gamma -1)\\
\label{eq5}
{\bf{\nabla}}\cdot {\bf{A}}=0.
\end{eqnarray}
Here $\gamma=1/\sqrt{1-{\bf{v}}^2/c^2}$ is the relativistic factor, $N$ is
the electron density, $N_0$ is the equilibrium density, $-e$ and $m$  the
electron charge and mass respectively, $\bf{A}$ is the electromagnetic
vector potential, $\varphi$ is the electrostatic scalar potential  and
$\psi$ is a scalar function which  expresses the electron canonical
momentum. Details on the derivation of this model can be found in
\cite{chen93}. Eqs.(\ref{eq3}) and (\ref{eq4}) imply that we are assuming
vortex-free motion of the electrons. Taking the divergence of
Eq.(\ref{eq1}) and using Eqs.(\ref{eq2}) and (\ref{eq5}) we find that the
charge conservation law
\beq \label{conservation}
\fr{\de N}{\de t} + {\bf{\nabla}} \cdot (N {\bf{v}}) =0
\eeq
is automatically satisfied, i.e., the total charge is conserved. However, when dealing
with necessarily simplified models describing the stationary regime in presence of electron
cavitation phenomena, the condition of
plasma quasineutrality is not obviously conserved, \cite{sun,borisov,kurki}. This point
must be carefully discussed when constructing new solutions and it will lead to the
breaking of the Hamiltonian model, thus allowing for multi-filament structures.

The assumption made on the pulse duration implies that the electron fluid
has time to approach a quasi-steady state, \cite{sprangle}. Therefore, it is
interesting to describe what kind of stationary state the system will
reach, neglecting any transient phenomena. This leads to further
simplification, because we can neglect the time dependence in
Eq.(\ref{eq4}) and, since $\bf{v}$ before the passage of the laser pulse
must be zero, it follows from Eqs.(\ref{eq3}) and (\ref{eq4}) that $\psi=0$.

In order to single out the fast optical time scale we adopt the slowly
varying envelope approximation, factorizing the normalized vector potential
as:
\beq
\fr{e {\bf{A}}}{m c^2}=a_{\perp}({\bf{r}}_{\perp}){\bf{r}}_{\perp} \exp
{[i(h z - \omega t)]} + c.c.
\eeq
Assuming the paraxial approximation $k_{\perp} << h$, where $k_{\perp}$ is
the transverse component of the laser wavenumber and h is the propagation constant,
the parallel component of
the vector potential is negligible if compared to the transverse ones
and the incident radiation may be assumed to be circularly
polarized without loss of generality.
In what follows, we will drop the
subscript denoting the perpendicular component of the various quantities.
The resulting system of equations, after a few algebraic manipulations, is:
\begin{eqnarray} \label{vectorpotential}
\nabla^2 a + \left( 1 - \fr{\alpha n}{\g}\right)a=0\\
\label{poisson}
\nabla^2 \phi  = \alpha (n -1)\\
\label{motion}
\phi = \g -1 \;\;\; \textrm{if and only if} \; n \neq 0\\
\label{gamma}
\g = \sqrt{1 + a^2}
\end{eqnarray}
where
\beq
\alpha= \fr{n_0}{1 - h^2/k^2}
\eeq
$k = \omega/c$ is the vacuum wavenumber and we have introduced the following normalization: $n_0 =N_0 /N_{cr}$ with
$N_{cr}=m
\omega^2/(4 \pi e^2)$, $n=N/N_0$, $\phi=e \varphi/(m c^2)$,
$\vec{r}=k\sqrt{1 - h^2/k^2}
\vec{r}_{\perp}$.

\section{Common Analysis}\label{common}

Let us consider a two dimensional geometry for a plasma extending in the
$z$ direction, i.e. along the laser propagation direction. We will restrict ourselves
to a 1D transverse model in order to emphasize the main features of multi-filament structures.
In
Eqs.(\ref{vectorpotential})-(\ref{gamma}) the propagation constant $h$ plays
the role of a free parameter which, together with the background plasma density $n_o$,
defines what kind of
filament structures can be realized as a final state of the self-focusing
evolution. In reality, it would depend on several parameters and factors as
the laser power, the geometrical configuration (the  angle of focusing, for instance) and
on the prehistory of the process.\\

The self-channeling we are interested in
is realized only  when $\alpha > 1$, i.e., for underdense
plasmas when $1 >h/k >\sqrt{1-n_0}$ and for overdense plasmas with $n_0 >1$ when $h/k <1$.

\begin{figure}
\begin{center}
\includegraphics[scale=0.8]{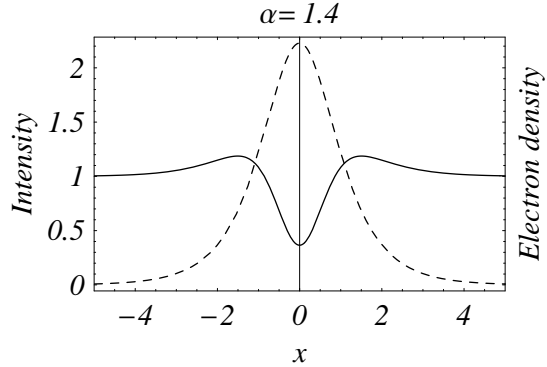}
\end{center}
\caption{{Plasma-field structures (dashed line) and electron density
distribution (continuous line) for the case of a single filament in an underdense plasma,
for a fixed value of $\alpha$ less then 1.5.
In this case $\alpha=1.4$.}}
\label{simple}
\end{figure}

\begin{figure}
\begin{center}
\includegraphics[scale=0.6875]{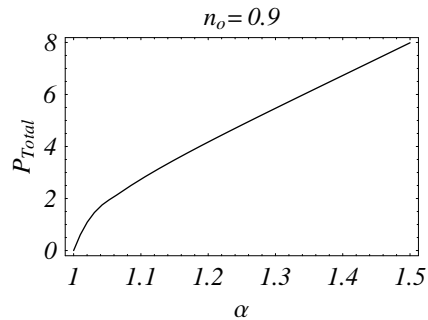}
\end{center}
\caption{{ The power needed to generate a single filament structure is shown
 as a function of the parameter $\alpha$ for $\alpha$ less then 1.5.
All  quantities are dimensionless.}}
\label{simpletwo}
\end{figure}

The complete mathematical analogy between the present model and the one introduced in
\cite{threshold} suggests that our plasma will react to the laser action with the
formation of regions depleted of electrons, where the laser electromagnetic
radiation is trapped, a consequence of the
well known phenomenon of electron cavitation and channeling. Electrons tend to be
expelled from the focal spot by the laser ponderomotive force and, at the same time
under such extreme conditions, they acquire relativistic quiver velocities. These
effects both contribute to a self-induced modification of the radial profile of the
refractive index and a consequent nonlinear trapping of the laser
radiation in finite plasma regions. This modification is the basic mechanism in the
optical guiding of laser pulses in plasmas. It is possible to give an exact analytical description of the asymptotic stationary
plasma-field structures generated in the transverse direction for different values
of $\alpha$. As we will see, these structures consist of one or more channels,
depending on the corresponding incident power. The most delicate point in the analysis,
will be the determination of
such structures complying with global charge conservation.\\
Eqs.
(\ref{vectorpotential})-(\ref{gamma}) were analyzed in detail in
\cite{kurki} and also in \cite{marburger} with respect to both underdense
and  overdense plasmas and solutions were found in the form of continuous functions.
Fundamental to those analysis is the Hamiltonian structure of the set of equations
(\ref{vectorpotential})-(\ref{gamma}), which reads
\beq \label{hamiltonian}
{\mathcal{H}}_E=\frac{1}{2 (1 + a^2)}a^{\prime 2}-\frac{1}{2}(2 \alpha
\sqrt{1 + a^2}-a^2)
\eeq
where the prime denotes the derivative with respect to the transverse coordinate
$x$. As $n(x) \rightarrow 1$ and both $a(x)$ and $a^\prime (x)$  vanish for
$x\rightarrow \infty$, the integral of motion equals
\beq \label{integral}
{\mathcal{H}}_E={\mathcal{H}}_{E0} \equiv -\alpha.
\eeq
It follows that there is an exact soliton-like analytical solution given by
\beq \label{sol}
a(x)= \fr{A_m \cosh
\left[|\varepsilon_0|^{1/2}(x-x^{(0)})\right]}{\alpha\cosh^2\left[|\varepsilon_0
|^{1/2}(x-x^{(0)})\right]-|\varepsilon_0|}
\eeq
where $\varepsilon_0 =1-\alpha $ and the parameter $x^{(0)}$ defines the
peak position of the function (\ref{sol}) which is given by $ A_m=2[\alpha
(\alpha-1)]^{1/2}$. Once this solution is known, we also have a description
for the electron density through Poisson's equation (\ref{poisson}) and the
equation of motion (\ref{motion}):
\beq \label{density}
n=3 ( 1+ a^2) + 2 \fr{\sqrt{1 + a^2}}{\alpha} \left({\mathcal{H}}_E - a^2
\right)
\eeq
The minimum electron density in a cavity is given by $n_{min}=1 - 4 (\alpha
-1)^2$, which implies that, for $\alpha > 1.5$, this solution leads to the
unphysical result of a negative electron density. Therefore, if $\alpha \leq 1.5$, i.e.,
for propagation constants lying in the interval $\sqrt{1-2n_0 /3} \leq h/k
> \sqrt{1-n_0}$, we have solutions of
 expressed by the continuous functions (\ref{sol})and (\ref{density}).\\
%
%
%
The field and density structures and the corresponding power, related to the propagation
constant, are presented in Fig. \ref{simple} and \ref{simpletwo}.


It should be emphasized that, for $\alpha \leq 1.5$, since the system is fully
described by the Hamiltonian (\ref{hamiltonian}), there are no other structures except
this single-filament ones.

\begin{figure}
\begin{center}
\includegraphics[scale=0.8]{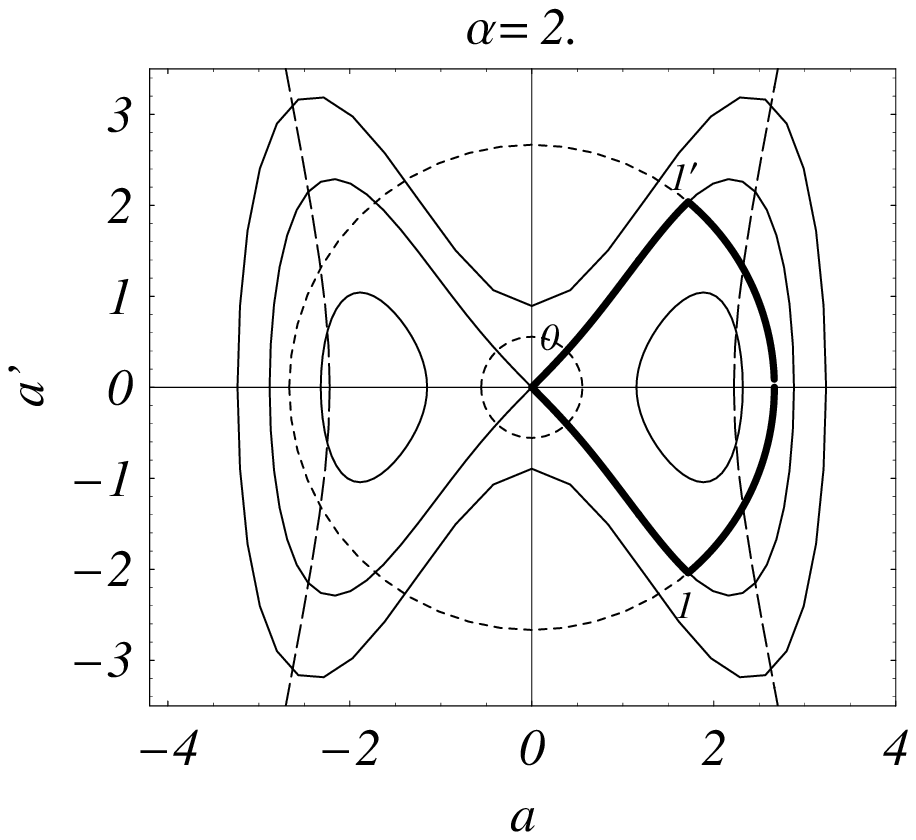}
\end{center}
\caption{{ Phase portrait for a system that develops a single-filament structure.
The thick line represents the trajectory followed by the system,
starting from the separatrix at $a=0, a^{\prime}=0$, then moving on along the vacuum trajectory to
finally come back to the starting point along the separatrix again which represents the symmetrical
plasma region. The corresponding plasma-field structures are illustrated in
Fig.\ref{singlefilament}.}}
\label{phase}
\end{figure}

\begin{figure}
\begin{center}
\includegraphics[scale=0.8]{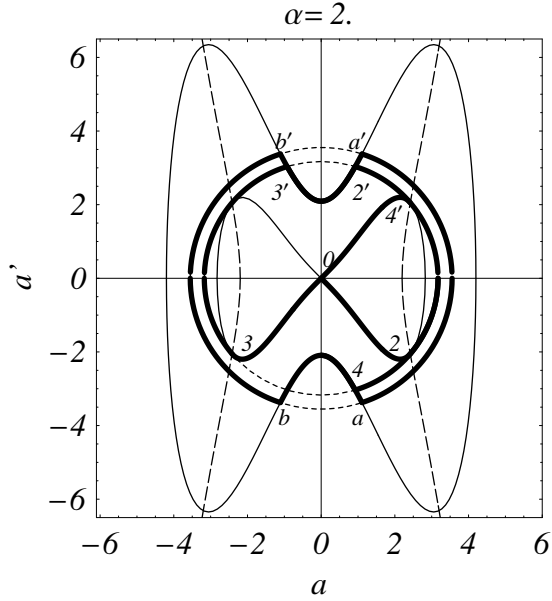}
\end{center}
\caption{{ Phase portrait for a more complicated case with multiple filaments.
The thick line represents the trajectory in the case of four filaments.
Solid lines represent trajectories relative to electron layers while dashed lines
are
relative to depletion regions and the large dashed line separates regions
with positive and negative electron density, as follows from Eq. (\ref{density}).}}
\label{phasetwo}
\end{figure}

The important question is what will happen for higher incident
powers or, in  other words, for $\alpha>1.5$?
 The procedure for constructing a solution followed in
\cite{borisov,kurki,feit96}, which consisted in assuming the electron density to
vanish within the interval where the solution for the density (\ref{density}) is negative,
led  to  non conservation of the global charge. However, what is happening is that
the ponderomotive force is pushing electrons away from the central axis, while the
force due to charge separation acts in the opposite direction. Thus, when an equilibrium is
reached, we have the
formation of a stationary structure consisting of a channel emptied of its electrons.
This means that the global structure of
the solution consists of two parts, the first one described by the Hamiltonian
(\ref{hamiltonian}) while the second, describing the depletion regions where the electron
density vanishes, has the typical vacuum Hamiltonian:
\beq \label{vacuum}
{\mathcal{H}}_V=
\fr{1}{2}(a^{\prime 2}+a^2).
\eeq
In Fig. \ref{phase}  the phase portrait
of the full system is presented for the single filament case with $\alpha=2$ while
in Fig. \ref{phasetwo} the same phase portrait is shown for
a more complicated multi-filament case, with $\alpha=2$.


The continuous soliton-like solution described by (\ref{sol}) corresponds here to the
separatrix trajectory and its starting and final point is $a=0, a^{\prime}=0$. As
pointed out, this solution breaks down for higher values of $\alpha$ and we indicate
on the phase portrait the curve
beyond which the electron density (\ref{density}) formally becomes  negative.\\
Beyond the limit curve for the electron density
 we have to introduce  the ''vacuum'' part of the
solution. Our system has left the separatrix and has started to move along
the vacuum one.
The  boundary position up to which the electrons are displaced is
determined by the equilibrium condition between the two forces acting on them, as
described by the equation of motion
\beq \label{equilibrium}
\phi^{\prime}=\gamma^{\prime}
\eeq
and by the conservation of the total charge,  which
means that, in order to conserve the total charge, the boundary positions can now be
determined by inserting the equilibrium condition (\ref{equilibrium}) into Poisson's
equation and integrating it.\\

\section{SINGLE-FILAMENT SOLUTIONS}

\begin{figure}
\includegraphics[scale=0.8]{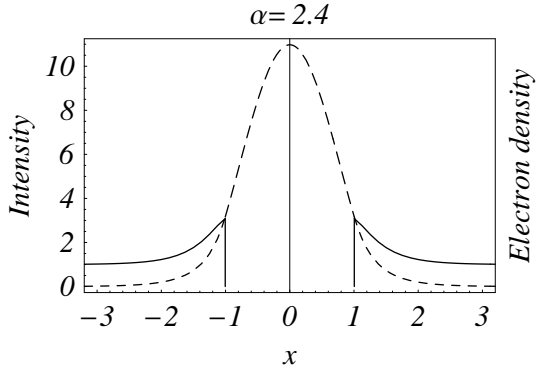}
\caption{{Plasma-field structures (dashed line) and electron density
distribution (continuous line) for the case of a single channel
for a fixed value of $\alpha=2$.
All  quantities are dimensionless.}}
\label{singlefilament}
\end{figure}

Let us consider a localized solution with one peak for the intensity. Its structure is defined by
the closed trajectory $(0-1-1^{\prime}-0)$ shown in Fig. \ref{phase}(a). We
will treat this kind of solution as a single-filament solution.\\
Integrating
Poisson's equation over the whole interval
we obtain
\beq
x_d=- \fr{1}{\alpha}\fr{a_d a_d'}{\sqrt{1 + a_d^2}}
\eeq
where $a_d$ is the field amplitude at the boundary position $x_d$. At the
same time, we have to match the field in the vacuum channel
\beq
a(x)=A_{V} \cos x
\eeq
and its first derivative to the field and its first derivative in the
plasma region, that is,
\beq
A_{V}^2=a_d^2 + a_d'^2
\eeq
and
\beq
x_d=- \arctan \left( \fr{a_d'}{a_d} \right)
\eeq
Given the integral of motion ${\mathcal{H}}_E=-\alpha$, from the two
equations for $x_d$ we obtain a transcendental equation for the boundary
amplitude $a_d$
\beq \label{transc}
\tan \left( \fr{a_d[2\alpha(\sqrt{1 + a_d^2}-1)-a_d^2]^{1/2}}{\alpha}
\right)=\\
\fr{\sqrt{1 + a_d^2}}{a_d}[2\alpha(\sqrt{1 + a_d^2}-1)-a_d^2]^{1/2}
\eeq
which can be solved numerically, so that now we know everything about the structures generated in
this case, see Fig. \ref{singlefilament}. It is important to be careful when solving
this equation since it has multiple solutions but we have to choose
the first
which
satisfies the condition of charge conservation.

\begin{figure}
\includegraphics[scale=0.8]{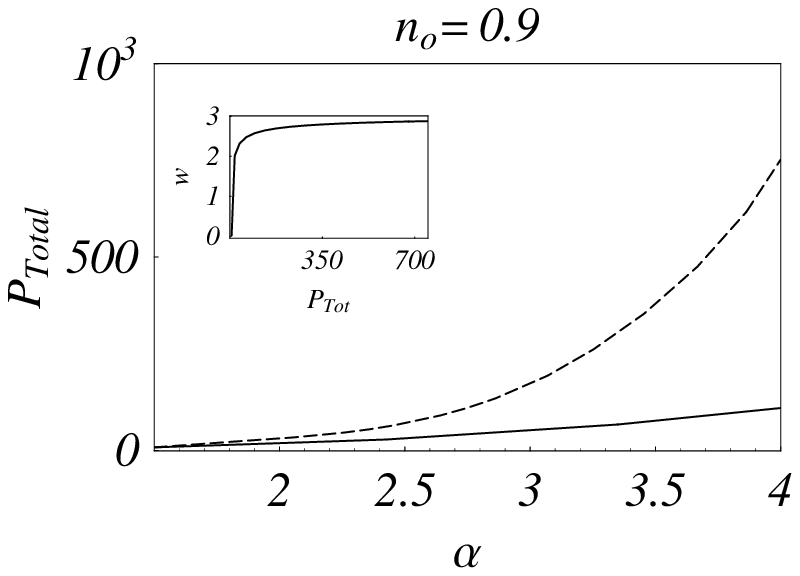}
\caption{{Total power (continuous line) and channel width $w=2 x_d$ (box) for the single filament case.
For comparison, the dashed line
shows the total power calculated according to the commonly used model,
without taking into account global charge
conservation.}}
\label{fig2}
\end{figure}

The calculation of the total power for this single
channel configuration as a function of $\alpha$ is straightforward:
$P=\int_{-\infty}^{+ \infty} a^2 (x) dx$, and the result is presented in
Fig.\ref{fig2}. For comparison, we present here also the total power
calculated within the previous model, when the boundary position $x_d$ was
assumed to be the one where the electron density vanished. In this case, the
total charge being not conserved, there was an excess of positive charge which led to
a much higher power required in order to overcome the restoring force due to this
charge excess. Consequently, the power needed to generate such structures was
overestimated. Besides, it is interesting to
see how, for increasing values of $\alpha$ and consequently increasing
values of the required power, the width of the central vacuum channel
becomes larger but, after the initial rapid growth, its increase becomes
slower, see the box in Fig.\ref{fig2}.

\section{MULTI-FILAMENT STRUCTURES}

It is evident from our analysis  that, due to the requirements of
global charge conservation and to the symmetry imposed with respect to the
$z$-axis, for a fixed $\alpha$ the single filament configuration and the power necessary to
generate this structure are uniquely determined. If this power is exceeded,
the incident electromagnetic radiation is strong enough to spread along the
transverse direction of the plasma channel over a distance larger than in
the previous case, but still finite. For a sufficiently strong power,
the final stationary state will present a multiple channel structure which
can again be analytically described. The power necessary for generating
each of these structures can be exactly calculated as well. As we will show, this
allows for the definition of a threshold power for the generation of multi-filament
structures.

\begin{figure}
\includegraphics[scale=0.8]{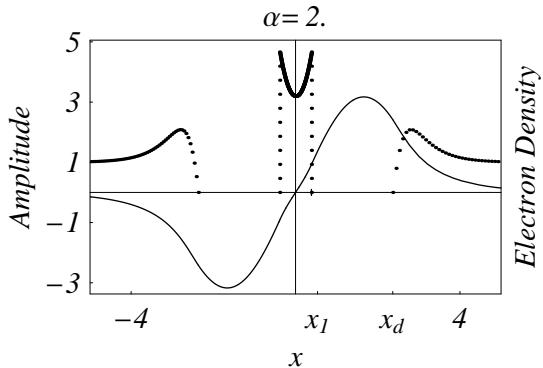}
\caption{{Double-channel structure for a plasma with a fixed $\alpha = 2$
and maximum
possible intensity $a_d^2$ at the last boundary (continuous line) and
electron density distribution (dotted line). All quantities are dimensionless.}}
\label{double}
\end{figure}

Let us start with the case of an intensity distribution with a minimum on
the symmetry axis. An example of the trajectory in the phase space for the double channel case is
given in  Fig. \ref{phase}(b), indicated as $(0-2-2^{\prime}-3^{\prime}-3-0)$.
 To construct the field structure we can start from
inside the plasma region at $x \rightarrow +\infty$, where we know the
integral of motion and the expression for the decaying field and the
electron density, see  Eqs. (\ref{sol}) and (\ref{density}) respectively.
When we come to the depletion region there is a certain freedom in the
choice of the boundary amplitude $a_d$, as we are going backwards from the
last plasma region towards the central axis. The only requirement for $a_d$
is that the electron density must not be negative, therefore we can fix the
boundary amplitude (and therefore the boundary position $x_d$ as well)
to any value up to a maximum for which the density at
the boundary vanishes. This means that, for a fixed $\alpha$ the two peak solution is
not unique and there is a certain power range for generating such a structure.
For a given $a_d$ the field in the vacuum region,
\beq
a(x)=A_V \cos (x-\varphi),
\eeq
is completely determined from the matching conditions, but now the vacuum
channel extends from the $x_d$ to a certain $x_1$ (see Fig. \ref{double})
which is to be determined taking into account charge conservation.

In order to construct a structure with only one degree of complexity more
than for the single channel, we stop at the next plasma layer, which will
be centered on the symmetry axis. An analytical expression for the field in
this central plasma layer can be derived by solving the equation for
the vector potential. Now the solution is not
localized as before, therefore the boundary conditions and the Hamiltonian
${\mathcal{H}}_E={\mathcal{H}}_{E1} > -\alpha$ are not known. The solution
is expressed in terms of two-parameter elliptic functions as
\begin{equation} \label{elliptic}
a(x)=\left\{
\begin{array}{l}
\frac{2q {\rm cn}[\varepsilon_1^{1/2}(x-x^{(1)})]}{2+[(q^2+1)^{1/2}-1]{\rm
sn}^2[\varepsilon_1^{1/2}(x-x^{(1)})]},\: -\alpha<{\mathcal{H}}_{E1}<\alpha, \\
\frac{2\bar{q}{\rm
sn}[((\varepsilon_1+1)^2-\alpha^2)^{1/2}(x-x^{(1)})/2]}{\bar{q}^2-{\rm
sn}^2[((\varepsilon_1+1)^2-\alpha^2)^{1/2}(x-x^{(1)})/2]},\:
{\mathcal{H}}_{E1}>\alpha
\end{array} \right.
\end{equation}
where $\varepsilon_1=(\alpha^2+1+2{\mathcal{H}}_{E1})^{1/2}$,
$q=[(\varepsilon_1+\alpha)^2-1]^{1/2}$ and
$\bar{q}=[(\varepsilon_1+\alpha+1)/(\varepsilon_1+\alpha-1)]^{1/2}$, while
$k=[(\alpha^2-(\varepsilon_1-1)^2)/4\varepsilon_1]^{1/2}$ and
$\bar{k}=[(\varepsilon_1-1)^2-\alpha^2)/(\varepsilon_1+1)^2-\alpha^2)]^{1/2}$ are
the moduli of the elliptic integrals of the first kind respectively for
the two cases. These solutions were presented in \cite{jetp} for the problem of self-induced
transparency of an overdense plasma.
Imposing the conservation of the total charge by integrating Poisson's
equation from $x=0$ to $x=\infty$ with the equilibrium condition defined by
the equation of motion, see Eq.(\ref{equilibrium}), we obtain a
transcendental equation for the quantity $\xi =x_1 - x_d$:
\begin{equation}\label{transcendental}
\xi=g(\xi)-g(0)
\end{equation}
where
\begin{equation}
g(\xi)=\frac{A_V^2\sin[2(\xi+\xi_0)]}{2\alpha[1+A_V^2\cos^{2}(\xi+\xi_0)]^{1/2}}
,
\; \xi_0=-\arctan\left(\frac{a^{\prime}_d}{a_d}\right)
\end{equation}
The solution of this equation gives a complete description of the vacuum
layer since $x_d$ is already determined. A
necessary condition for this equation to have a non-trivial solution is
that  $g^\prime (\xi=-\xi_0)>1$, i.e.
\begin{equation} \label{gprime}
g^\prime (\xi=-\xi_0)=\frac{A_V^2}{\alpha(1+A_V^2)^{1/2}}>1.
\end{equation}
which cannot be satisfied unless $\alpha \geq 1.5$. This leads to the
conclusion that if $\alpha < 1.5$, that is, if the propagation constant of
the wave vector $h$ is not large enough, our system will never reach a
stationary state and it will display only a dynamical behaviour with the
electromagnetic perturbation propagating along the transverse direction.
Otherwise, we can numerically calculate how the minimum boundary intensity,
such that Eq. (\ref{transcendental}) still has a solution,
depends
on $\alpha$ and also find a solution of
this equation and calculate $x_1$. To finally obtain a complete description of
the central plasma layer we only have to apply the boundary conditions at
$x_1$ to determine the parameters that are still unknown,
${\mathcal{H}}_{E1}$ and $x^{(1)}$, while the symmetry axis is determined
as the symmetry axis of the elliptic function. In Fig.\ref{double} a
double-channel structure is shown for fixed $\alpha$ and for maximum
amplitude at the last boundary (so that the electron density at this
boundary vanishes). It should be noticed that, for a fixed value of the
boundary amplitude $a_d$, the width of the vacuum channels and the peak
intensity in these channels increase with $\alpha$. Furthermore, the
maximum possible boundary amplitude itself is an increasing function of
$\alpha$ and, for any given value of this parameter, such a maximum
amplitude determines the maximum power we can deliver to the plasma in
order to generate a double-channel structure. Exceeding this maximum power will force
the system to generate a structure with one more filament and therefore we can talk about
a threshold power for the generation of multi-filament structures.

\begin{figure}
\includegraphics[scale=0.8]{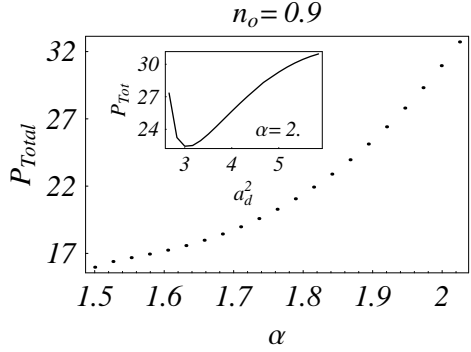}
\caption{{Total power required to generate a double channel structure
versus $\alpha$ for the case of maximum possible intensity at the last boundary and (box)
as a function of the boundary intensity  for fixed $\alpha=2$.}}
\label{powerdouble}
\end{figure}

In Fig.\ref{powerdouble} we show $P_{Total}$ calculated
for varying $\alpha$ and for $a_d$ fixed to its maximum possible
value. In the box it is shown instead how the total power varies with the intensity
at the last boundary.

The apparently anomalous behaviour
for  low boundary intensities is due to the fact that
the left branch corresponds to a new kind of two-filament solution whose phase portrait
field structures are presented
in Fig. \ref{nonzero}.\\

As can be seen following the trajectory ($0-1-2-3-2{\prime}-1{\prime}-0$), in this case
the field amplitude, once it leaves the separatrix, never crosses the zero point
until it reaches the separatrix again. For those periodic trajectories lying inside
the separatrix, the Hamiltonian ${\mathcal{H}}_{E1}$ is less than $-\alpha$ and the
field solution for the electron layers assume a different form and
are now described in terms of elliptic
functions as
\begin{equation}
a(x) = \fr{(\alpha - \varepsilon_1 -1) \{1- (\varepsilon_1^{1/2}/\varepsilon_2) {\rm
sn}^2 [\varepsilon_2 (x - x^{(1)})] \}^{1/2}}{1 - 2 (\varepsilon_1/\alpha +
\varepsilon_1-1) {\rm sn}^2 [\varepsilon_2 (x - x^{(1)})]}
\eeq
where $\varepsilon_1=(\alpha^2 +1+\mathcal{H}_{E1})^{1/2}$, $\varepsilon_2=[\alpha^2
- (\varepsilon_1-1)^2]/2$ and the module of the elliptic integral of the first kind is
$k=\varepsilon_1^{1/2}/\varepsilon_2$. The procedure to define the electron cavitation
boundaries is the same as the one followed previously to build the structures
presented in Fig.\ref{double}.

\begin{figure}
\begin{center}
\includegraphics[scale=.9]{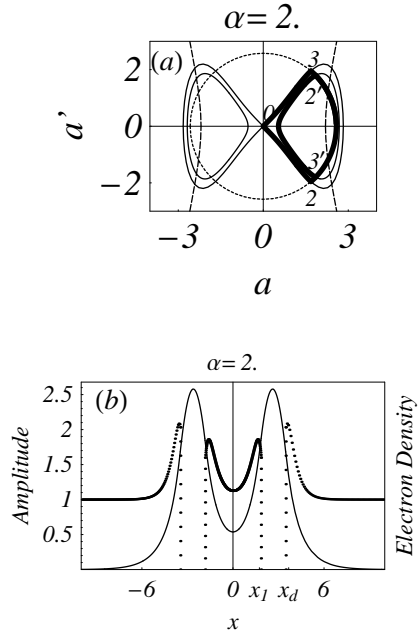}
\end{center}
\caption{{(a) Phase portrait for a system that develops a double-filament structure.
In this case the field amplitude never vanishes apart from $\pm \infty$ on the
separatrix.
The corresponding plasma-field structures are illustrated in part
(b).}}
\label{nonzero}
\end{figure}

The solution we have constructed and the corresponding choice of a closed one-cycle
trajectory in the phase space is not unique.
 We can pass a depletion region not only at the point
$3\prime$ to form a one cycle trajectory,  but also at the point  $b^{\prime}$
in order to create a periodic trajectory as  $b^{\prime}-b-a-a^{\prime}$, see Figs. \ref{phase}(b)
and \ref{nonzero}(a).
Following
this trajectory means that we will have a new
structure with new channels and plasma layers, which were not present in the double channel
structure previously described.\\

We would like to underline the fact that the periodic trajectory shown in Fig.
\ref{phase} corresponds to a particular configuration as the points  $b^{\prime}, a^{\prime}, a,
b$ are related by a complete symmetry. It is possible to see what this means by
looking at the field structures described by such a trajectory (see Fig. \ref{four}):
The  central channels are completely symmetric, at each boundary we have the same
intensity.
It is again the request of global charge conservation which leads, by integrating
Poisson's equation, to a transcendental
equation for the field at the boundary of the new plasma layer:
\beq
\tan \left( \fr{a_d[2{\mathcal{H}}_{E1}+2\alpha\sqrt{1 +
a_d^2}-a_d^2]^{1/2}}{\alpha} \right)=\fr{\sqrt{1 +
a_d^2}}{a_d}[2{\mathcal{H}}_{E1}+2\alpha\sqrt{1 + a_d^2}-a_d^2]^{1/2}
\eeq
This equation is similar to Eq.(\ref{transc}) but now the Hamiltonian value
${\mathcal{H}}_E={\mathcal{H}}_{E1}$ is the one defined for the new electron layer.
In the case of a single-peak field distribution, this equation had a unique solution,
consequently we can add to the two-filament distribution whole periodic cycles, i.e.,
even numbers of filaments by jumping to consecutive vacuum trajectories at the point
$b\prime$ or repeating the same vacuum trajectories,
as seen in Fig.\ref{four} where a six-filament structure is presented, which
corresponds to the
trajectory  $(0-2-2^{\prime}-b^{\prime}-a\prime-a-b-3^{\prime}-3-0)$, with two cycles
along the periodic trajectory $b^{\prime}-a\prime-a-b$. Again, it is
important to notice that this structure, completely symmetric, is peculiar of a planar
geometry.

\begin{figure}
\includegraphics[scale=0.8]{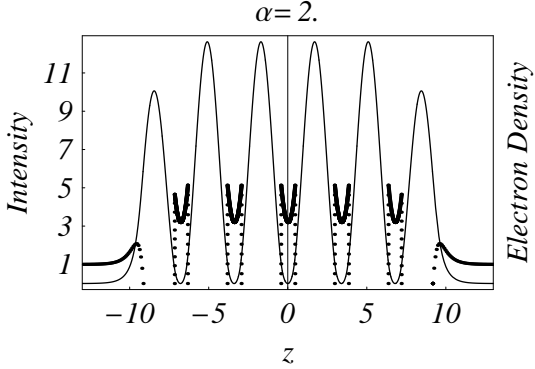}
\caption{{Six-channel structure for a plasma with  fixed $\alpha=2$
and maximum possible intensity at the last boundary (continuous line) and
electron density distribution (dotted line). All quantities are dimensionless.}}
\label{four}
\end{figure}

We can also add to the single filament configuration an  odd number of filaments, considering trajectories
corresponding to a certain number of cycles plus half a cycle. Consider, for example,
the trajectory
$(0-2-2^{\prime}-b^{\prime}-b-4-4^{\prime}-0)$, where the points
$4, 4^{\prime}$ are symmetrical to $2^{\prime}, 2$ respectively, which
corresponds to a three-filament structure. The result is shown, for a fixed
value of $\alpha$ in Fig.\ref{triple}.

\begin{figure}
\includegraphics[scale=0.8]{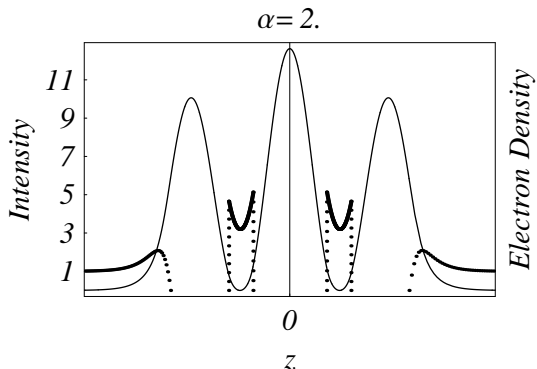}
\caption{{Three-channel structure for a plasma with  fixed $\alpha=2$
and maximum possible intensity at the last boundary (continuous line) and
electron density distribution (dotted line). All quantities are dimensionless.}}
\label{triple}
\end{figure}

Finally, in Fig.\ref{allpower}, we present in the same graphic the calculated maximum powers as functions of
$\alpha$ for  three of the different cases we have analyzed, single, double and triple
channel.

\begin{figure}
\includegraphics[scale=0.8]{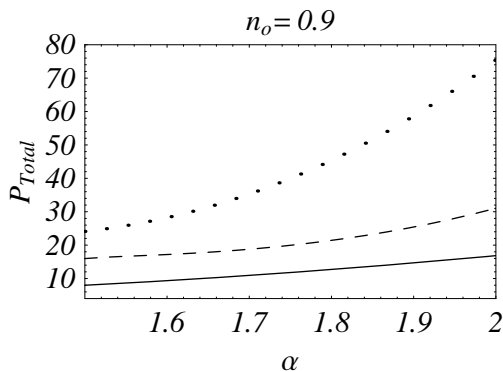}
\caption{{Total power as a function of $\alpha$ for the three different cases: single channel (continuous line),
double channel (dashed line), triple channel (dotted line). In each case, the
amplitude at the last boundary was chosen as the maximum possible one.}}
\label{allpower}
\end{figure}

The same procedure that we have described, may also be applied to the case presented
in Fig.\ref{nonzero} where an integer number of filaments can be added since a full
cycle occurs within a half-space of the phase portrait.

Therefore, by using this procedure, we can construct multi-filament solutions that
exist only for $\alpha>1.5$ and represent plasma channels with electron
cavitation. They differ from each other because of the laser power transported along
these channels and, as there is a minimum laser power required for exciting
such structures, we can define the power thresholds for creating
non-single filament structures.

\section{CONCLUSIONS}\label{conclusions}

In conclusion, we have presented an exact analysis of self-channeling structures
generated as a consequence of the relativistic self-focusing due to the interaction of
ultra-intense laser radiation with an underdense plasma.
In this analysis, the
plasma quasineutrality condition is accurately taken into consideration and
this  quantitatively affects some results on channeling laser power.
The essential point of the analysis is  that it allows us to prove the multi-filament nature of
 the relativistic self-focusing and to calculate the threshold power  for
exciting multi-filament structures. Such a result is not achievable in media with
a local nonlinearity, like the Kerr one, because the governing equation has an overall Hamiltonian
structure. In the case we have analyzed, each electron
cavitation channel corresponds to a certain part of the trajectory
followed by the system in the phase space, each part with its own
 Hamiltonian value, as shown in Fig.\ref{phase}. Concerning the definition of a
 threshold power, it is interesting to notice how filament structures with regions
 depleted of their electrons can be generated by the interaction if the parameter
 $\alpha$ is greater than 1.5. For $\alpha \leq 1.5$ we have only single filament
 field structures, with no depletion regions. As soon as $\alpha$ exceeds 1.5 there
 are plasma regions that are emptied of their electrons and the number of filaments
 thus generated increases with increasing incident power. We can therefore define the
 maximum power incident on a plasma with $\alpha=1.5$ as the real threshold power for
 generating non-single filament structures (see Fig.\ref{simple}(b)).
 The same construction procedure
 followed for the
single-filament solution can then be easily  extended to a more realistic
axisymmetrical case to obtain the analogous of Figs. (5) and (7) , while a real 2D transverse
approach is needed  for the multi-filament structure problem, especially for the case presented in
Fig.(8) where a number of equal filaments have been added to the fundamental
structure.

\section{ACKNOWLEDGMENTS}

This work was partly supported by INTAS (grant no. 96-339).
One of the authors (F.C.) would like to acknowledge support from the
European Community (TMR program) under the contract ERBFMBICT972428.

\end{document}